\documentstyle[epsf,fleqn,aps]{revtex}

\def\Section#1{}

\def\Sz{S^z}

\def\beq{\begin{equation}}
\def\eeq{\end{equation}}
\def\bea{\begin{eqnarray}}
\def\eea{\end{eqnarray}}

\begin{document}
\tightenlines
\title{Instabilities in Luttinger liquids}
\author{D.C.\ Cabra$^{1,2}$ and J.E.\ Drut$^1$}
\address{$^1$Departamento de F\'{\i}sica, Universidad Nacional de la Plata,
C.C.\ 67, (1900) La Plata, Argentina.\\ $^2$ Facultad de
Ingenier\'{\i}a, Universidad Nacional de Lomas de Zamora,\\ Cno.\
de Cintura y Juan XXIII, (1832), Lomas de Zamora, Argentina.}

\maketitle

\begin{abstract}
We discuss the appearance of magnetic and charge instabilities,
named respectively metamagnetism (MM) and phase separation (PS),
in systems which can be described by a perturbed Luttinger liquid.
We argue that such instabilities can be associated with the
vanishing of the effective Fermi velocity $v$, which in some cases
coincides with a divergence of the effective Luttinger parameter
$K$. We analyze in particular an $XXZ$ chain with
next-nearest-neighbor interactions in different limits where MM
shows up and an extended Hubbard model where in turn, PS occurs.
Qualitative agreement with previous studies is found.

\vspace{10 pt}


\vspace{-12 pt}

\end{abstract}

\vskip2pc


\section{Introduction}

The study of instabilities in low dimensional strongly correlated
electron systems has received much attention in the last few
years. One of the main reasons is that a charge instability
phenomenon (phase separation (PS)) often shows up in the vicinity
of the superconducting transition in cuprates. In the case of
double exchange models for manganese oxides that exhibit the ``Colossal"
magnetoresistance effect \cite{MYD}, this charge
instability arises close to the transition to ferromagnetism and,
interestingly, finite size studies of both the two-dimensional
realistic model and its one-dimensional version display similar
features in this respect. The case of magnetic instabilities
(metamagnetism (MM)) has also received recent attention in
connection to the one dimensional antiferromagnetic (AF)
$XXZ$ spin chain with
next-nearest-neighbor interactions (NNN), where it was found that
MM arises in a finite region of the phase space \cite{GMK}.

Generically, at the point where charge or magnetic instabilities
occur, a divergent compressibility or magnetic susceptibility
arises. This divergence is in turn associated respectively with
the coexistence of two phases with different hole concentrations
or magnetizations.

The aim of this paper is to discuss a general way to determine
whether an instability could show up in a given one-dimensional
model, using Abelian bosonization. Our discussion applies to any
one-dimensional model describable as an integrable model plus
perturbations whose effect is to renormalize the Luttinger liquid
(LL) parameters, $K$ and $v$. The effect of irrelevant operators
is also discussed. Our discussion could be also relevant in the
study of certain 2D systems which can realize the so called
sliding Luttinger phase \cite{SLL}, since a divergence in the 1D
susceptibility leads to a singularity in the 2D one \cite{2Dchi}.
Generically, our approach provides a quick tool to study the
tendency of different perturbations to produce instabilities
in the above mentioned systems.

As a sample case for magnetic systems, we analyze the $XXZ$ chain
with NNN exchange, treating first this last interaction
perturbatively within a bosonization approach and find qualitative
agreement with the results obtained in \cite{GMK} in finite
systems (This problem was also studied in \cite{MM}). We also
study the opposite coupling limit, which we call hereafter
``zig-zag" limit, where the system can be reinterpreted as a
two-leg zig-zag ladder. In this case, we determine a region in the
parameter space where MM occurs.

To analyze the case of charge instabilities we consider the
charge sector of the
Hubbard model at incommensurate fillings, perturbed by a
nearest-neighbor density-density interaction $V$. In this case we
find that for small Coulomb repulsion $U$ there is a region where
the system phase separates which corresponds to negative values of
$V$, in agreement with previous studies \cite{PM}.

\section{General discussion}

We consider a generic situation in which the low energy degrees of
freedom (which could correspond to either charge or spin
variables) are described by a Tomonaga-Luttinger liquid in the
unperturbed case. The corresponding Hamiltonian once interactions
are taken into account will generically be of the form (modulo
irrelevant terms)

\beq
H_0 + \alpha (\partial_x \phi)^2 + \beta (\partial_x
\tilde{\phi})^2 + \lambda (\partial_x \phi) \ ,
\label{maineq}
\eeq
where $H_0$ corresponds to the Tomonaga-Luttinger Hamiltonian
\beq
H_0 = {1 \over 2} \int dx \left( v K (\partial_x \tilde\phi)^2
+ {v \over K} (\partial_x \phi)^2 \right) \, .
\label{LL}
\eeq

The first two terms can be readily absorbed into a redefinition of
the LL parameters $K\rightarrow K_{eff}$ and $v\rightarrow
v_{eff}$, while the third one changes the chemical potential in
the case of charge variables and the magnetic field in the spin
case.

Under the above mentioned conditions the compressibility for the
charge modes, described by an effective LL with parameters $K_c$
and $v_c$ can be shown to be given by \cite{FK}

\beq
\kappa \propto \frac{K_c}{v_c} \ ,
\label{comp}
\eeq
and similarly, the magnetic susceptibility for the spin modes
described by an effective LL with parameters $K_s$ and $v_s$ is
given by

\beq
\chi \propto \frac{K_{s}}{v_{s}} \ .
\label{susc}
\eeq

One readily observes that a divergence in these quantities arises
either when $K_{(c,s)}^{-1}$ or $v_{(c,s)}$ vanish. It should be
pointed out that these two things could happen simultaneously, but
this is not always the case and hence PS or MM instabilities
are to be identified with the vanishing of the effective velocity.
We argue that this is true, provided
that the quantities $Kv$ and $K/v$ remain positive definite and
irrelevant perturbations do not change the large scale behavior. In
Ref.\ \cite{japos} an attempt to characterize these instabilities
in one dimensional systems exhibiting a universal character
described by a Tomonaga-Luttinger model was made. In this paper
the authors identified the divergence of the above mentioned
thermodynamic quantities with a divergence in the so called
Luttinger parameter $K$. Although this interpretation led to a
consistent analysis for the cases studied in \cite{japos}, we
argue that a more general criterion consists in identifying the
instability regions with those where the velocity vanishes. This
last statement can be understood as follows: by analogy with the
unperturbed $XXZ$ chain in a magnetic field, we see that when the
Fermi velocity goes to zero we approach a ground state of FM
nature (and $K$ does not necessarily diverge, though this happens
for the particular case of the $XXZ$ chain for $\Delta = -1$ and zero field
\cite{japos}). Then the magnetic susceptibility diverges as
$1/v_{eff}$ and, if this happens before reaching saturation, the
magnetization curve as a function of the applied magnetic field
presents a jump. Once we have shown that there is a jump in the
magnetization curve, for some critical value of the applied
magnetic field, $h_c$, we can conclude that two different values
of the magnetization will coexist at this point \cite{Elbio}.

In the next two Sections we study two sample cases where this
general discussion applies, the $XXZ$ chain with NNN interactions
where MM has been shown to occur \cite{GMK}, \cite{MM} and the
extended Hubbard model with nearest-neighbour interactions $V$,
where PS appears for negative $V$ \cite{PM}.

\section{The $XXZ$ AF chain with NNN interactions}

The Hamiltonian of the $XXZ$ AF with NNN interactions in a
magnetic field is given by

\bea H_{XXZ} = J \sum_{i=1}^N \left( S^x_iS^x_{i+1} +
S^y_iS^y_{i+1} + \Delta S^z_iS^z_{i+1}\right) + \nonumber \\ J'
\sum_{i=1}^N \left( S^x_iS^x_{i+2} + S^y_iS^y_{i+2} + \Delta
S^z_iS^z_{i+2}\right) - h \sum_{i=1}^N \Sz_i . \label{xxz} \eea

The large scale behaviour of the $XXZ$ chain can be described by a
$U(1)$ free boson theory with Hamiltonian (\ref{LL}). The field
$\phi^i$ and its dual $\tilde{\phi}^i$ are given by the sum and
difference of the light-cone components, respectively. The constant
$K$ governs the conformal dimensions of the bosonic vertex
operators and can be obtained exactly from the Bethe Ansatz
solution of the $XXZ$ chain (see e.g.\ \cite{CHP} for a detailed
summary). We have $K=1$ for the $SU(2)$ symmetric case ($\Delta =
1$) and is related to the radius $R$ of \cite{CHP} by $K^{-1} = 2
\pi R^2$. In (\ref{LL}) $v$ corresponds to the Fermi velocity of
the fundamental excitations of the system.

In terms of these fields, the spin operators read

\bea
S_x^z
&\sim& {1 \over \sqrt{2\pi}} \partial_x \phi + a : \cos(2 k_F x +
\sqrt{2 \pi} \phi): + \frac{\langle M \rangle}{2} \, , \label{sz}
\\S_x^{\pm} &\sim& (-1)^x :e^{\pm i\sqrt{2\pi} \tilde{\phi}} \left(b
\cos(2 k_F x + \sqrt{2 \pi} \phi) + c \right) : \, ,
\label{s+}
\eea
where the colons denote normal ordering with respect to the
groundstate with  magnetization $\langle M \rangle$. The Fermi
momentum $k_F$ is related to the magnetization of the chain as
$k_F = (1-\langle M \rangle )\pi/2$. The effect of an $XXZ$
anisotropy and/or the external magnetic field is then to modify
the scaling dimensions of the physical fields through $K$ and the
commensurability properties of the spin operators, as can be seen
from  (\ref{sz}), (\ref{s+}). The constants $a$, $b$ and $c$ were
numerically computed in the case of zero magnetic field \cite{HF}
(see also \cite{LA}).

In what follows we study both the weak coupling ($J'/J \ll 1$) and
the zig-zag ($J'/J \gg 1$) limits.

\vspace{.5cm}

\noindent {\bf i)  $J'/J \ll 1$ limit}

\vspace{.2cm}

Using (\ref{sz}) and (\ref{s+}) the NNN interaction term in the
bosonic language reads

\beq
H_{NNN}= \alpha \int dx \left(g_1
\left(\partial_x \phi\right)^2 + g_2 \left(\partial_x
\tilde{\phi}\right)^2  + \lambda ~ \partial_x \phi
+ \lambda' ~ \cos (\sqrt{8\pi} \phi)
\right),
\label{NNN}
\eeq
where $\alpha = J'/J$ and $g_{1,2}$, $\lambda$ and  $\lambda'$
depend on $\Delta$ and the non-universal constants $a$, $b$ and
$c$ as

\beq
g_1 = \Delta \left(\frac{1}{2\pi} - 2\pi a^2 \cos(4k_F)
\right) + 2\pi b^2 \cos(4k_F),
\label{g1}
\eeq

\beq
g_2 = -4\pi
\left(c^2 - \frac{b^2}{2} \cos(4k_F) \right) ,
\label{g2}
\eeq
and
\beq
\lambda=\frac{1}{\sqrt{2\pi}} \alpha \Delta \langle M \rangle ,
\ \ \ \ \ \ \ \lambda'=\alpha ~ b ~ c  .
\label{lambda}
\eeq
%
%

The first two terms in (\ref{NNN}) have the effect of renormalizing both the
compactification radius and the Fermi velocity in the following
way

\beq
v_{eff}^2= \left(v/K + 2 \alpha g_1\right)\left(vK + 2
\alpha g_2\right)  ,\ \ \ \ \ \ K_{eff}^2=\frac{vK + 2\alpha
g_2}{v/K + 2\alpha g_1},
\label{rentot}
\eeq

Now we can make contact with the analysis of Ref.\ \cite{GMK}: the
MM region is identified within our approach as the set of phase
space points in which the susceptibility diverges for $0 < M <
M_{sat}$.

One immediately observes from (\ref{rentot}) that, for zero
magnetic field and without taking into account the effect of the
$\lambda'$ perturbation,
both $v_{eff}$ and $K_{eff}^{-1}$ vanish
simultaneously and that this happens when $\left(v/K + 2 \alpha
g_1\right)=0$ (see Fig.\ 1). However, this situation can change for two reasons:
first, an extra term arises from the NNN perturbation, which
renormalizes the external magnetic field $h$ and hence changes
both $v_{eff}$ and $K_{eff}$ through the Bethe Ansatz equations,
generically in a way which removes the above mentioned
simultaneity. Second, if one goes beyond the zero loop order, both
$K_{eff}$ and $v_{eff}$ will renormalize in a different way due to
the $\lambda'$ term
\cite{G}. We will discuss this issue again in the context of
charge instabilities in the next Section. In particular, the
boundary between the MM and the FM phases is obtained as the set
of points in which $v_{eff}$ and/or $K_{eff}^{-1}$ vanish/es for
$h=0$, and that between the MM and the AF phases as the set of
points in which $v_{eff}\neq 0$ and $K_{eff}^{-1}\neq 0$ for all
values of the magnetization $\langle M\rangle < M_{sat}$.

Let us focus on the MM-FM boundary, which is easily obtained using
(\ref{susc}) and (\ref{rentot}). In this case, we can use the
numerical values for the non-universal constants $a$, $b$ and $c$
appearing in (\ref{sz}) and (\ref{s+}) obtained in \cite{HF}. The
boundary obtained in this way agrees qualitatively with that
obtained in \cite{GMK} (see Fig.\ 1). The lack of quantitative
agreement is presumably due to the perturbative treatment of the
NNN interactions, which could be improved by considering higher
loop contributions in a renormalization group analysis from the bosonization side
and due to finite size effects from the numerical one. However,
our main aim is to discuss the appearance of instabilities in a
generic and simple way.

One should be careful also about the regime of validity of this
approach, since due to the renormalization of the Luttinger
parameter $K_{eff}$, the scaling dimensions of the many discarded
irrelevant perturbations change, and nothing prevents one of these
to become relevant. In fact, by analyzing the scaling dimension of
the leading irrelevant $\lambda'$ perturbation, we observe that
it reaches the limiting value 2 at $\alpha_c =
\frac{v(K^2 - 1)}{2K(g_1 - g_2)}$, and hence our approach ceases
to be valid for $\alpha \ge \alpha_c$. This critical line
separates the massless regime from a massive one, and
our results compare qualitatively well with those obtained in
\cite{japos2} (dashed line in Fig.\ 1).
The study of the AF-MM boundary is in this case more involved due
to the appearance of the non-universal constants $a$, $b$ and $c$
in the bosonized operators which are not available for non-zero
magnetic field. (The estimation of the field dependence of these
constants for $0 \le \Delta < 1$ has been done in  \cite{HF2} but
its evaluation is more delicate on the ferromagnetic side $\Delta
< 0$ \cite{pc}.)

\begin{figure}
\hbox{ \epsfxsize=4.1in \epsffile{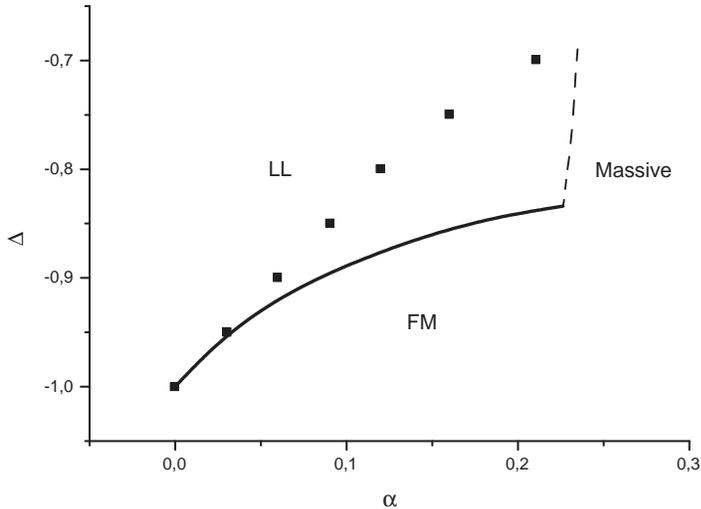}} \vspace{1cm}
\caption{Ground state phase diagram for the $XXZ$ chain with
next-nearest-neighbor interactions in the limit $J'/J \ll 1$
showing the transition lines between the Luttinger liquid regime
and the ferromagnetic one. Square dots are reproduced from [2] and
the solid black line corresponds to our theoretical prediction.
The dashed line indicates where the leading irrelevant operator
becomes marginal.}
\end{figure}

\vspace{.5cm}

\noindent {\bf ii) Zig-zag limit ($J'/J \gg 1$)}

\vspace{.2cm}

In this limit we reinterpret the Hamiltonian (\ref{xxz}) as a two
leg zig-zag ladder in the weak interchain coupling limit
\cite{CPKSR,WA,AS,NGE,Sor,nosxxz}. In this description one
represents each of the chains by one free compactified $U(1)$
boson. Thus, the whole ladder is represented by two bosons
$\phi_1$, $\phi_2$, each governed by an action given by
(\ref{LL}), plus the perturbative terms arising from the
interchain zig-zag coupling.

This system, in the presence of an external magnetic field, has
been studied in \cite{nosxxz}, where it was shown that for
non-zero magnetization a relevant interaction gives a mass to the
``relative" field $\phi_{rel} \equiv (\phi_1 - \phi_2)/2$, while
the ``diagonal" field, $\phi_{diag} \equiv (\phi_1 + \phi_2)/2$,
remains massless.

The effective Hamiltonian governing the large scale behavior of
$\phi_{diag}$ is then given by

\beq
H_{eff} = {1 \over 2} \int dx \left( v_{eff} K_{eff}
(\partial_x \tilde{\phi}_{diag})^2 + {v_{eff}\over K_{eff}}
(\partial_x \phi_{diag})^2 +\lambda \partial_x \phi_{diag}
\right) ,
\label{Hamdiag}
\eeq
where, calling $\beta = 1/\alpha$,
$\lambda = \beta \Delta \langle M\rangle /\sqrt{2\pi}$
and $v_{eff}$ and $K_{eff}$ are given by

\beq
v_{eff}^2= v^2\left(1 + \frac{\beta \Delta K}{\pi v}\right)
,\ \ \ \ \ \ K_{eff}^2=K^2\left(1 + \frac{\beta \Delta K}{\pi
v}\right)^{-1}.
\label{rendiag}
\eeq

There is however an extra term that mixes the fields $\phi_{rel}$
and $\phi_{diag}$

\beq
H_{mix} = - \lambda' \int dx \partial_x
\tilde{\phi}_{diag} \sin (\sqrt{4\pi}\tilde{\phi}_{rel}).
\label{mix}
\eeq
The effect of this term was studied in Ref.\ \cite{NGE} where it
was shown that it gives rise to a spin nematic phase close to
$\Delta =0$. The main point for our analysis is that the diagonal
field is still described by a LL, and we expect the same picture
to apply for $-1< \Delta\le 0$. For this reason, we do not take
into account this last term in what follows.

One then observes from  (\ref{rendiag}) that the non-universal
constants do not appear in the perturbing terms, and hence we can
obtain in this case both the FM-MM and MM-AF boundaries.
The results are presented in Fig.\ 2.

\begin{figure}
\hbox{%
\epsfxsize=4.1in
\epsffile{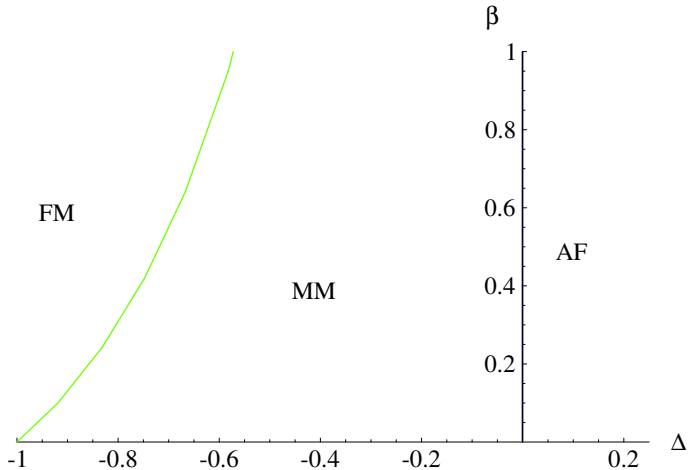}}
\vspace{1.5cm}
\caption{Qualitative ground state phase diagram for the $XXZ$
chain with next-nearest-neighbor
interactions in the limit $J'/J \gg 1$.}
\end{figure}

\section{Hubbard model with nearest neighbor interactions}

We consider now the Hubbard model, defined as
\beq H = -\frac{1}{2}\sum_{j,\alpha}{\left(\psi^\dagger_{j,\alpha}
    \psi_{j+1,\alpha} + h.c. \right)} +
    U\sum_j {n_{j,\uparrow} n_{j,\downarrow}} +
    \mu\sum_{j,\alpha} n_{j,\alpha},
    \label{HubHam}
\eeq
where \beq n_{j,\alpha} =
\psi^\dagger_{j,\alpha}\psi_{j,\alpha}.\eeq

In the absence of an external magnetic field it has been shown
that this model presents charge-spin separation. It has also been
shown (c.f. \cite{PencSolyom}) that the large scale behavior of
the spin and charge degrees of freedom can be described by
two decoupled boson field theories with dynamics governed by the
Tomonaga-Luttinger Hamiltonian. The parameters $K$ and $v$ can in
each case be exactly obtained for all values of $\mu$ and $U$ via
numerically solving the Bethe Ansatz equations in \cite{LiebWu}.
Approximate expressions for the velocity of the charge sector in
the small and large $U$ regimes are given in \cite{PencSolyom}.

The addition of a density-density interaction between nearest
neighbors leads to the so called Extended Hubbard model, whose
Hamiltonian is given by (\ref{HubHam}) plus the term

\beq
    \delta H = V\sum_{j}{n_j n_{j+1}},
    \label{PertHubHam}
\eeq
where

\beq
n_j = n_{j,\uparrow}+n_{j,\downarrow}.
\eeq

It is a simple matter to show that, modulo irrelevant operators,
the effect of the perturbation (\ref{PertHubHam}) is to
renormalize the parameters $\mu$ and $U$ as follows:

\beq
\mu \rightarrow \mu + V ,\ \ \ \ \ \  U \rightarrow U +
2V.
\eeq

Therefore, the low energy behavior of the charge sector of the
Extended Hubbard model is that of a Luttinger liquid with
parameters

\beq
K_{eff} = K(U + 2V,\mu + V), \ \ \ \ \ \ \  v_{eff} = v(U +
2V,\mu + V),
\eeq
where $K(U,\mu)$ y $v(U,\mu)$, are given by the exact Bethe Ansatz
solution of the non-Extended Hubbard model (\ref{HubHam}).
This allows for the determination of the divergences in the
compressibility $\kappa$, as given by (\ref{comp}), which we
identify with the zeros of the effective velocity.

We have only considered small values of $V$, since our approach is
a perturbative one. Within this restriction, we have found no
singularities, and therefore no PS, in the large $U$ regime. On
the other hand, regarding the small $U$,$V$ region, in which the
effective velocity is given by

\beq
    v_{eff} = v(U + 2V,\mu + V) \sim \frac{\pi}{2}\left(1 + \sqrt{1 + 2(\mu +
    V)}\right)(\mu + V) + \frac{U + 2V}{2 \pi} \ ,
\label{veffhub}
\eeq
we have found that the roots of the above expression present the
same qualitative behavior as that described in \cite{PM}. It
should be stressed that $K_{eff}$ remains finite in the region
where (\ref{veffhub}) vanishes.

\begin{figure}
\hbox{%
\epsfxsize=4.1in \epsffile{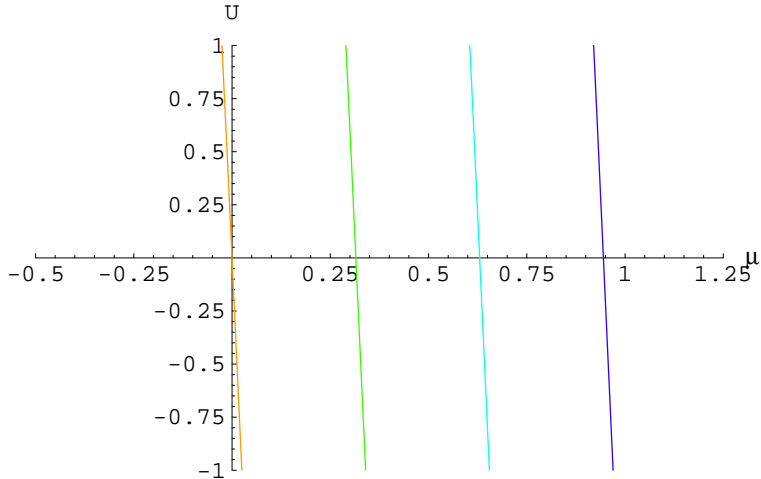}} \vspace{1.5cm}
\caption{Phase separation
lines in the extended Hubbard model for $V=0,\, -0.3,\,
-0.6,\, -0.9$ from left to right.}
\end{figure}

Following this approach, one could also study the appearance of
instabilities in the magnetic sector.

\section{Conclusions and discussion}

We have discussed the appearance of MM and PS in systems that can
be described as perturbed Luttinger liquids. Specifically, we have
studied the $XXZ$ spin chain with NNN interactions and the
extended Hubbard model with nearest-neighbor density-density
interactions. We have found FM, AF and MM regions in the zig-zag
limit of the $XXZ$-NNN spin chain. In the weak coupling limit,
qualitative agreement with previous results was found for the
MM-FM transition. We were not able to complete our search for MM
due to lack of specific numerical data, an issue to be discussed
elsewhere.  The instabilities were in all cases identified as the
roots of the inverse susceptibility (\ref{susc}).

Concerning the extended Hubbard model, according to our analysis,
it does not present PS in the large $U$ and small $V$ limit. On
the other hand, in the small $U$, $V$ limit instabilities do show
up, through the roots of the effective Fermi velocity, while the
effective Luttinger parameter remains finite.

We argue that all the instabilities studied are associated in
general with a vanishing effective velocity $v_{eff}$. As a matter
of fact, it is the roots of $v_{eff}$ that give a divergent
compressibility of the extended Hubbard model, though in the other
cases studied, this coincides with a divergence of the effective
Luttinger parameter $K_{eff}$. This should not be regarded as the
actual hallmark of the instabilities: although our treatment of
the $XXZ$-NNN spin chain gives at the same time $v_{eff} =
K_{eff}^{-1} = 0$, this is only true in the absence of an external
magnetic field and moreover ceases to be valid beyond the zero
loop order since both parameter renormalize in a different manner
\cite{G}. It should be stressed that our approach gives a simple
way to qualitatively analyze instabilities in generic charge and
magnetic systems provided they can be described as perturbed
Luttinger Liquids. Henceforth, it provides a quick tool to study
whether different perturbations could produce such
instabilities.

\vspace{.5cm}

\noindent {\it Acknowledgments:} It is a pleasure to acknowledge
fruitful discussions with M.\ Arlego, E.\ Dagotto, A.\ Furusaki,
M.D.\ Grynberg, T.\ Hikihara, A.\ Honecker and P.\ Pujol.  The
authors acknowledge partial financial support of CONICET and
Fundaci\'on Antorchas.


\end{document}